\documentclass[letterpaper, 10 pt, conference]{ieeeconf}
\IEEEoverridecommandlockouts                              
\usepackage{graphics} % for pdf, bitmapped graphics files
\usepackage{subfigure}
\usepackage{epsfig} % for postscript graphics files
\usepackage{gensymb}
\usepackage{textcomp}
\usepackage{amsmath} % assumes amsmath package installed
\usepackage{algorithm,algorithmicx,algpseudocode}
\title{\LARGE \bf
Evaluation of Automated Vehicles Encountering Pedestrians at Unsignalized Crossings
}

\author{Baiming Chen$^{1}$, Ding Zhao$^{2}$, Huei Peng$^{3}$% <-this % stops a space
\thanks{* This work is funded by the Mobility Transformation Center Denso Tailor Project at the University of Michigan with grant No. N020210. }% <-this % stops a space
\thanks{$^{1}$B. Chen is with the Department of Automotive Engineering, Tsinghua University, Beijing, China, 100084, and is currently a visiting scholar at the University of Michigan, Ann Arbor, MI 48109, USA.}%
\thanks{$^{2}$D. Zhao is with the University of Michigan Transportation Research Institute, Ann Arbor, MI 48109, USA  (correspondig author: {\tt\small zhaoding@umich.edu}).}%
\thanks{$^{3}$H. Peng is with the Department of Mechanical Engineering, University of Michigan, Ann Arbor, MI 48109, USA.}%
}

\begin{document}
\maketitle
\thispagestyle{empty}
\pagestyle{empty}
%%%%%%%%%%%%%%%%%%%%%%%%%%%%%%%%%%%%%%%%%%%%%%%%%%%%%%%%%%%%%%%%%%%%%%%%%%%%%%%%
\begin{abstract}
Interactions between vehicles and pedestrians have always been a major problem in traffic safety. Experienced human drivers are able to analyze the environment and choose driving strategies that will help them avoid crashes. What is not yet clear, however, is how automated vehicles will interact with pedestrians. This paper proposes a new method for evaluating the safety and feasibility of the driving strategy of automated vehicles when encountering unsignalized crossings. MobilEye\textsuperscript{\textregistered} sensors installed on buses in Ann Arbor, Michigan, collected data on 2,973 valid crossing events. A stochastic interaction model was then created using a multivariate Gaussian mixture model. This model allowed us to simulate the movements of pedestrians reacting to an oncoming vehicle when approaching unsignalized crossings, and to evaluate the passing strategies of automated vehicles. A simulation was then conducted to demonstrate the evaluation procedure.

\end{abstract}
%euroncap
%%%%%%%%%%%%%%%%%%%%%%%%%%%%%%%%%%%%%%%%%%%%%%%%%%%%%%%%%%%%%%%%%%%%%%%%%%%%%%%%
\section{Introduction}
Traffic safety has become an issue of growing concern in the U.S. According to the NHTSA, the country lost 35,092 people in crashes on U.S. roadways during 2015, up 7.2\% from 32,744 in 2014, the largest increase in nearly 50 years \cite{HighwayTrafficSafetyAdministration2014}. One approach to decreasing traffic deaths is to use automated vehicles, which do not suffer from challenges such as fatigue, distraction and drunk driving that humans might face. Researches have been conducted on the interaction between vehicles and vehicles \cite{Zhao2016n,Zhao2016m,Huang2016UsingScenario}. What is not yet known, however, is exactly how automated vehicles will interact with pedestrians. Ways to test automated vehicles must be found before the vehicles can be put into use. The most critical places with a high concentration of interactions are intersections, particularly unsignalized crossings, where the right of way is not clear. As it is difficult for automated vehicles to decide the appropriate strategy for passing through an unsignalized intersection with pedestrians crossing the street, this scenario is a good one to consider.

The first stage of the evaluation requires an analysis of pedestrian behaviors. Early studies on pedestrian movement focused on pedestrians that did not interact with vehicles. Three typical kinds of pedestrian models were developed: discrete cellular automata models \cite{Burstedde2001}, continuous force-based models \cite{Chraibi2009}, and macroscopic pedestrian stream models \cite{Xiong2011}.

Additional attributes of pedestrian behavior were studied when vehicles were introduced into the picture. For example, the walking speed of pedestrians might vary depending on several factors, one of which being where pedestrians crossed the street. Thus, Bennett et al. \cite{Bennett2001} found that the average speed of those crossing at mid-block was slower than those crossing at a signalized intersection. Additional factors significantly contributing to pedestrian walking speed (i.e., age, gender, group size and street width) were found by Tarawneh \cite{Tarawneh2001} in Jordan. Pedestrians 21-30 years of age were found to be the fastest; those over 65, the slowest. The author also found that pedestrians tended to walk faster after longer wait times.
%Fitzpatrick et al. \cite{Fitzpatrick2006AnotherSpeed} found that there is a statistical difference in walking speeds between older (older than 60 years) and younger (60 years and younger) pedestrians, the 15th percentage walking speed for young pedestrians is 1.15 m/s and the 15th percentage walking speed for older pedestrians is 0.92 m/s. 
Whether pedestrians were in a group or alone also had an impact. Yagil \cite{Yagil2000} found, for instance, that pedestrians were more likely to wait at an intersection if a group of pedestrians were already waiting there.

One approach to analyzing pedestrian behaviors is to apply the concept of gap acceptance. 
%At an unsignalized intersection, pedestrians must make a decision before crossing the road: Is the gap in the traffic stream large enough to allow for a safe crossing? 
When arriving at an unsignalized intersection, pedestrians must decide whether the gap in the traffic stream large enough for them to cross the street safely.
To describe the gap acceptance behavior of pedestrians, probit and binary models were developed by Sun et al. \cite{DazhiSun2002}. Regression analysis found the important factors for pedestrian gap acceptance to be gap size, number of pedestrians waiting, and pedestrian age. Schroeder \cite{Schroeder2008} improved gap acceptance models for unsignalized crossings by incorporating vehicle dynamics, pedestrian characteristics and concurrent events at the crosswalk. This model is difficult to use for the evaluation of automated vehicles, however, because information about the pedestrians (e.g., age and gender) cannot as yet be detected by the vehicles approaching the intersection.

The driving strategy of human drivers when encountering unsignalized crossings was studied. To describe driver yielding and pedestrian gap acceptance behaviors, empirical logit models were developed by Schroeder et al. \cite{Schroeder2014}. The author also proposed a driver dynamic model when faced with one crossing pedestrian. This model is introduced and used as the driving strategy of automated vehicles in the simulation section of this paper.
% * <zhaoding1014@gmail.com> 2016-12-26T15:26:30.694Z:
%
% > The driving strategy of human drivers when encountering unsignalized crossings was needed for the simulation.
%
% To do simulation is not the purpose. Just describe the literature of the AV control.
%
% ^.
A key component of the driving strategy is the braking behavior of drivers when a pedestrian comes out from the sidewalk to the road. For example, Suzuki et al. \cite{KeisukeSuzuki2016} concluded that the timing of the braking operation is approximately the same in terms of TTC (Time to Collision), though the velocity of the subject vehicles and passing pedestrians are different.

%Schroeder et al. \cite{Schroeder2014} studied vehicle-pedestrian interaction at unsignalized pedestrian crossings and developed empirical logit models describing driver yielding and pedestrian gap acceptance behaviors. The author also proposed a driver dynamic model when facing one crossing pedestrian with linear regression analysis. The data were recorded by a video camera at the crosswalk, so the types of vehicles in the recorded passing events are not the same, which may influence the effectiveness of the data. However, data collected by in-car sensors will not have this problem since all the event data were recorded by the same vehicle. 

%However, there are currently no driver models at unsignalized crossings when facing multiple pedestrians.
%However, too many parameters were used for the regression which increased the difficulty of evaluation

%However the number of samples were small and the drivers were practiced to do the same experiment for many times.
%Akagi et al. \cite{Dqg2015} proposed the risk potential model that describes the braking behavior when a driver is passing unsignalized intersections. However, the influence of pedestrians' crossings was not studied.
Before automated vehicles are put into use, they must be evaluated for their potential interaction with pedestrians.
The driving strategy of automated vehicles should be neither too aggressive (endangering traffic safety) nor too conservative (leading to a waste of time and traffic congestion). Current experimental approaches are limited, however. The European New Car Assessment Programme (Euro NCAP) established a test procedure for AEB VRU systems, the automatic emergency braking systems that specifically look for and react to pedestrians  \cite{Euro-NCAP2015}. The purpose of that test differs from the one in this paper, however. Our goal is to simulate the movement of multiple pedestrians and test the automated vehicles' passing strategy, while the procedure proposed by Euro NCAP was designed to test the effectiveness of the braking system with the sudden appearance of only one pedestrian. More importantly, if an automated vehicle's driving strategy is too conservative, it  may still pass the NCAP test, but leads to traffic congestion and road rage in reality.
%but it will not pass the test proposed in this paper.

%In this paper, we did a series of work including data collection, interaction modeling, and designing of an evaluation method for testing the safety and feasibility of automated vehicles' passing strategy at unsignalized crossings.

\begin{figure}[h!]
  \centering
  \includegraphics[width=0.5\textwidth]{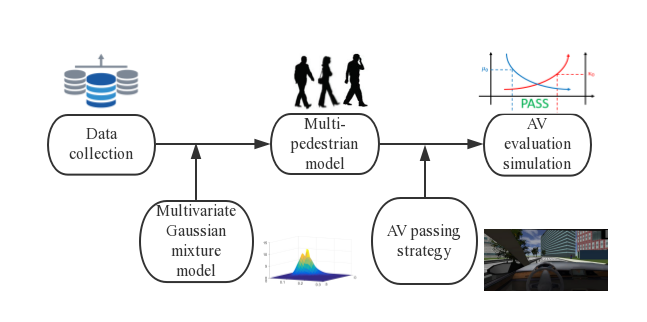}
	\caption{Procedures for designing the evaluation method}
  \label{fig:mesh1}
\end{figure}

The procedures used in this paper are shown in Fig. \ref{fig:mesh1}. A stochastic interaction model was first developed based on naturalistic traffic data. Then a method for evaluating the automated vehicles was proposed. Finally, the test procedure was demonstrated in simulation.

%The ideas and algorithms in this study are clear and in common use.

% \begin{figure}[h!]
%   \centering
%   \includegraphics[width=0.5\textwidth]{flow.png}
%   \caption{Process of this study}
% \end{figure}

The main contributions of this paper are:
\begin{itemize}
\item Naturalistic data of 2,973 passing events encountering pedestrians at unsignalized crossings were recorded by in-car sensors. To the best of our knowledge, this is the largest data set describing vehicle-pedestrian interactions that has ever been collected.
\item A stochastic interaction model was created by fitting a bounded multivariate Gaussian mixture model to traffic data.
\item A new method for evaluating automated vehicles at unsignalized crossings was proposed.
\end{itemize}

% Firstly, we collected passing event data by the Mobileye sensors installed on the Blue Buses in Ann Arbor, which can automatically distinguish pedestrians and record their relative positions to the bus. We selected the data that took place at the pedestrian crossing near the Central Campus Transition Center (CCTC) in Ann Arbor. The number of passing events we collected is 2973 at this crossing. More details will be discussed in Section I.

% Secondly, we fit Gaussian mixture distribution to the collected data with a 4-variable, 10-component model. With this model, we can simulate the subject vehicle's movement and the crossing pedestrian's walking speed under the given conditions. The method we used will be demonstrated in Section II.

% Lastly, we designed an experiment to evaluate the passing strategy of automated vehicles at unsignalized crossings. PreScan software was used for the simulation of the interactions between vehicles and pedestrians. Existing driver models were tested and the results were shown. See Section III for more information.

\section{Data Collection}

\begin{figure}[h!]
  \centering
  \includegraphics[width=0.5\textwidth]{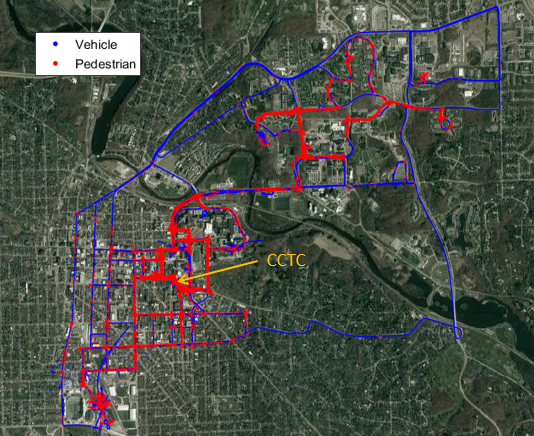}
  \caption{Routes of the instrumented buses and detected pedestrians in Ann Arbor, Michigan. The  Central Campus Transit Center (CCTC) has a high pedestrian density. The figure is created based on Google Maps.}
  \label{fig:mesh1111}
\end{figure}

The first step in being able to simulate the movements of pedestrians and vehicles was to collect naturalistic data. In this study, data on 2,973 passing events encountering pedestrians at unsignalized crossings were collected by in-car sensors.
The devices used to collect pedestrian behavior were MobilEye\textsuperscript{\textregistered} installed on university buses in Ann Arbor, Michigan. The university's 12 bus routes carry approximately 7.2 million passengers per year. 
%More information on the university buses can be found at https://pts.umich.edu/transit/getting-around.php. 
Three of the university buses participating in the SPMD (Safety Pilot Model Deployment) program \cite{bezzina2014safety} were equipped with MobilEye\textsuperscript{\textregistered} which
%MobilEye\textsuperscript{\textregistered}'s vision-based sensors 
can distinguish pedestrians and provide their relative positions to the instrumented vehicle. In addition, u-Box GPS/IMU unit installed on the university buses can provide the geographical location. Vehicle velocity and yaw rate were recorded from CAN bus. The routes of the instrumented buses and pedestrians detected by MobilEye\textsuperscript{\textregistered} are shown in Fig. \ref{fig:mesh1111}.

% Variables collected with sensors that we use in this study contains:
% \begin{itemize}
% \item\textit{\textbf{Range (m)} ------ Longitudinal position of an object.}

% \item\textit{\textbf{Transversal (m)} ------ Lateral position of the object.}
% \end{itemize}

% They can help determine the pedestrian's relative position. See Fig. \ref{fig:mesh2}.

% \begin{figure}[h!]
%   \centering
%   \includegraphics[width=0.5\textwidth]{rangetrans.png}
%   \caption{Definition of Range and Transversal}
%   \label{fig:mesh2}
% \end{figure}

% We can also get the vehicle $Speed (m/s)$ from the CAN signal and location ($Latitude, Longitude$) from GPS device.

\begin{figure}[h!]
  \centering
  \includegraphics[width=0.5\textwidth]{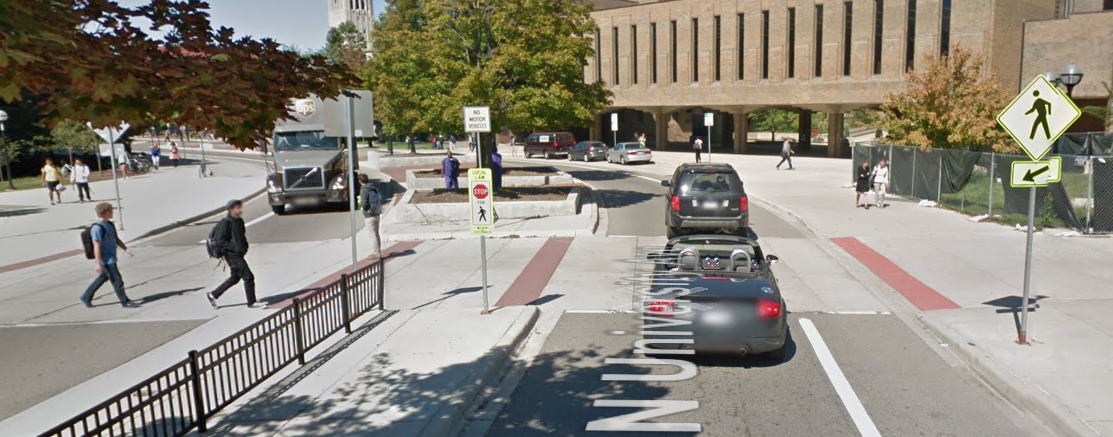}
  \caption{Crossing path environment (based on Google Maps)}
  \label{fig:mesh22}
\end{figure}

The unsignalized crossing chosen for collecting data is in Ann Arbor, Michigan, near the University of Michigan's Central Campus Transit Center (CCTC), where the pedestrian density is high (Fig. \ref{fig:mesh1111}). The geographic location is $42.278415^{\circ}$, $-83.735580^{\circ}$. A picture of this crossing path is shown in Fig. \ref{fig:mesh22}, where 2,973 valid passing events were collected.

%With the data collected, we can draw every passing event and observe the trajectories of the subject vehicles and pedestrians. With a coordinate transformation, we can draw the events in a Cartesian coordinate system. One example is shown in Figure \ref{fig:mesh3}. The color bar indicates the time of the samples.

\begin{figure}[ht!]
  \centering
  \includegraphics[width=0.5\textwidth]{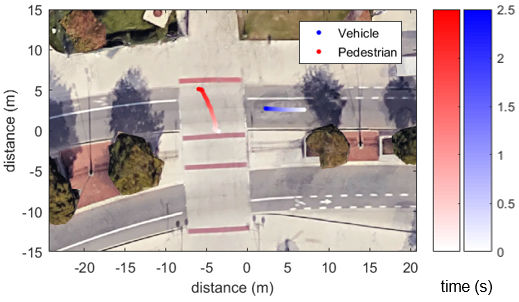}
  \caption{Sample of passing events (based on Google Maps)}
  \label{fig:mesh3}
\end{figure}

The GPS system recorded the latitude and longitude of the instrument vehicles, while the MobilEye\textsuperscript{\textregistered} recorded the relative positions of pedestrians. Using the data collected, each passing event was drawn in a Cartesian Coordinate system, so that the trajectories of the vehicles and pedestrians could be observed. One example is shown in Fig. \ref{fig:mesh3}. The color bar indicates the time of the samples.

\section{Interaction Model}
A multivariate Gaussian mixture model is created based on naturalistic data to simulate the movement of vehicles and pedestrians when encountering unsignalized crossings. This stochastic model demonstrates how vehicles and pedestrians will interact with each other. This section shows the procedures used to develop this model and how to use it for the simulation.
\subsection{Variables}

\begin{figure}[h!]
  \centering
  \includegraphics[width=0.5\textwidth]{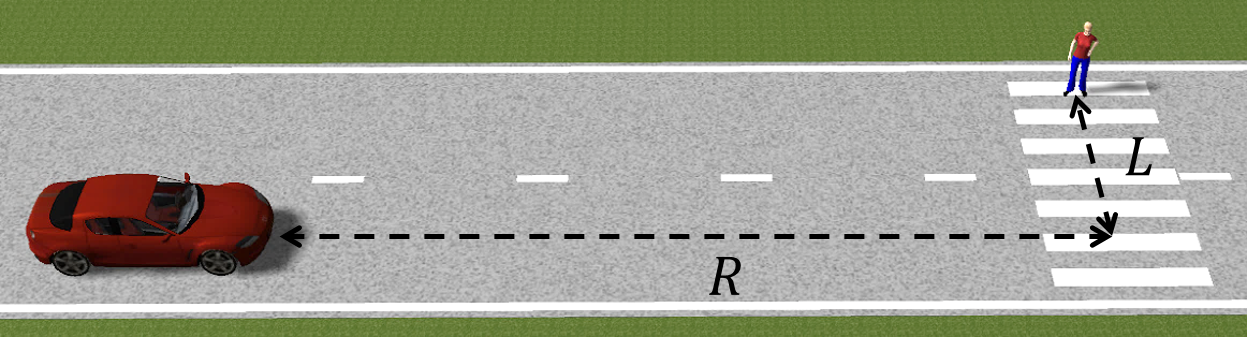}
  \caption{Definition of $R$ (longitudinal distance) and $L$ (lateral distance)}
  \label{fig:mesh222}
\end{figure}

Appropriate variables must first be selected to develop a multivariate Gaussian mixture model. We simplified the passing scenario by assuming that the trajectories of the vehicle and the pedestrian are two perpendicular straight lines. Then, longitudinal distance ($R$) and lateral distance ($L$) are used to describe the relative position of a pedestrian (Fig. \ref{fig:mesh222}).

% The list of parameters used in this section is shown in Table \ref{table1}.
% \begin{table}[!ht]
% \caption{List of parameters in the interaction model}
% \begin{center}
% \begin{tabular}{ccc}
% \hline
% \hline
% symbol & unit & value \\
% \hline
% $R$ & m & longitudinal position of the pedestrian \\
% %\hline
% $L$ & m & lateral position of the pedestrian \\
% %\hline
% $v$ & m/s & speed of the vehicle \\
% %\hline
% $v_p$ & m/s &walking speed of the pedestrian \\
% %\hline
% $TTC$ & s &time to collision \\
% %\hline
% $T_{Adv}$ & s &time advantage \\
% \hline
% \end{tabular}
% \end{center}
% \label{table1}
% \end{table}

%Among these parameters, longitudinal position ($R$) and lateral position ($L$) describe the relative position of a pedestrian (Fig. \ref{fig:mesh222}).

% * <zhaoding1014@gmail.com> 2016-12-26T04:50:43.526Z:
%
% Try to use prescan to draw the pedestrian
%
% ^.
% Time-to-Collision ($TTC$) is the time required for the vehicle to reach the crossing path at its present speed:
% \begin{equation}
% \label{ttc}
% TTC = R/v.
% \end{equation}

Four variables are required to define the state of the passing scenario: the reciprocal of the longitudinal distance ($R^{-1}$), vehicle speed ($v$), pedestrian walking speed ($v_p$) and the reciprocal of the Time Advantage ($T_{Adv}^{-1}$). Time Advantage is an indicator used to describe situations where two road users pass a common spatial zone, but at different times, thus avoiding a collision \cite{Laureshyn2010b}. The definition of Time Advantage is the time between the first road user leaving the common spatial zone and the second arriving. The mathematical definition is:
\begin{equation}
T_{Adv} = |TTC - L/v_p|.
\label{tadv}
\end{equation}
The reciprocals are used for statistical convenience. Among the four variables, $R^{-1}$ and $v$ can be obtained by the instrumented vehicle directly, $T_{Adv}^{-1}$ can be calculated by (\ref{tadv}), and $v_p$ can be calculated by the lateral distance ($L$) data.

% \begin{itemize}
% \item\textit{\textbf{Range (r)\textsuperscript{-1}}} --- Reciprocal of the longitudinal distance between the subject vehicle and pedestrian.

% \item\textit{\textbf{Vehicle Speed (v\textsubscript{0})}} --- Approaching speed of the vehicle.

% \item\textit{\textbf{Pedestrian Speed (v\textsubscript{p})}} --- Walking speed of the pedestrian to pass the crossing.

% \item\textit{\textbf{Collision avoidance distance (CAD)\textsuperscript{-1}}} --- Reciprocal of the distance between the subject vehicle and pedestrian after TTC assuming they maintain the consistent velocity. See Fig. \ref{fig:mesh4}.
% \end{itemize}

% \begin{figure}[!h]
%   \centering
%   \subfigure[Time = 0]{\includegraphics[width=0.23\textwidth]{rangetrans1.png}}
%   \hfill
%   \subfigure[Time = TTC]{\includegraphics[width=0.23\textwidth]{cad.png}}
  
%   \caption{Demonstration of CAD}
%   \label{fig:mesh4}
% \end{figure}

% Fig. \ref{fig:mesh4} describes the definition of CAD. Assume $time = 0$ in Fig. \ref{fig:mesh4}(a), and from this moment the vehicle and pedestrian maintain their velocity. So after $TTC = Range/VehicleSpeed$ the vehicle will reach the crossing and there will be a distance between the vehicle and pedestrian, and we name this distance CAD, see Fig. \ref{fig:mesh4}(b). The mathematical expression is 
% \begin{equation}
% CAD = v_p * TTC - Transversal
% \end{equation}

\subsection{Algorithms}
The probability density function ($p.d.f.$) of a Gaussian mixture model is 
\begin{equation}
f(y;\Theta)=\sum_{k=1}^{K}\pi_k f_k(y;\theta_k),
\end{equation}
where $\pi_k$ are weights of components, $f_k$ are density functions of each component, $\theta_k$ are parameters to decide each component. 
%Each observation is assumed to be from one of the $K$ components. 
The distribution of each component is a multivariate Gaussian distribution with mean $\mu_k$ and covariance $\Sigma_k$.

Algorithms are required to fit Gaussian mixture models to the traffic data. Here, since all the variables (i.e., $R^{-1}$, $v$, $v_p$, $T_{Adv}^{-1}$) are positive and have boundaries, a truncated Gaussian mixture model will better fit the data. Lee et al. \cite{Lee2012} developed for fitting multivariate Gaussian mixture models to data that is truncated. By applying these algorithms to the data, parameters $\pi_k$, $\mu_k$ and $\Sigma_k$ of the multivariate Gaussian mixture model can be calculated.

\begin{figure}[!h]
  \centering
  \subfigure[Change of BIC]{\includegraphics[width=0.23\textwidth]{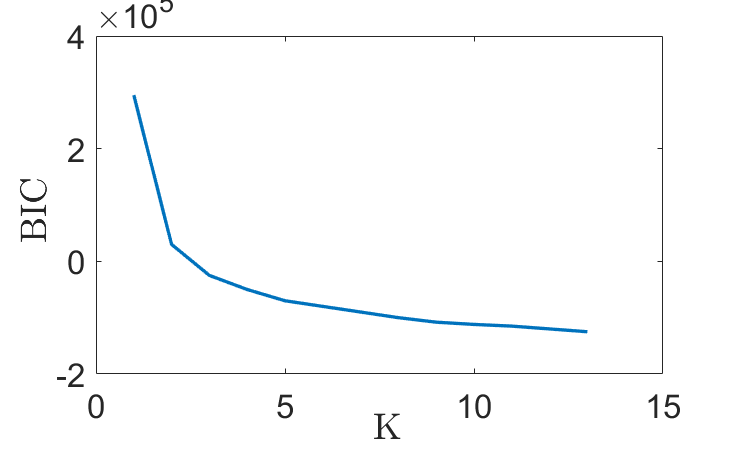}}
  \hfill
  \subfigure[Change Rate of BIC]{\includegraphics[width=0.23\textwidth]{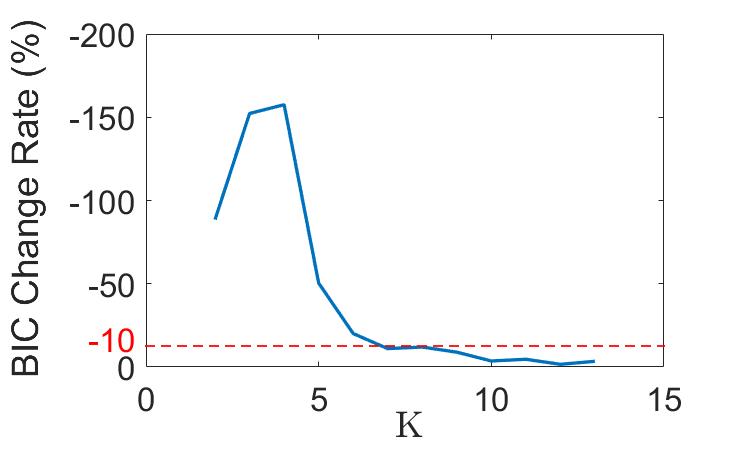}}
  \caption{Bayesian Information Criterion test of the interaction models with different number of components ($K$). The BIC continues to decrease with an increase in number of components, with little change in rate (less than 10\%) when the number of components is larger than 10.}
  \label{fig:mesh5}
\end{figure}

The number of components ($K$) of the multivariate Gaussian mixture model will influence how well the data fit, with a higher $K$  helping to improve the precision of the modeling. However, it might also lead to greater computing time, as well as statistical errors such as singular covariance matrices which endanger the stability of the model. Thus, an appropriate $K$ must be selected for the Gaussian mixture model. The Bayesian information criterion (BIC) provides a means for model selection. It can help estimate the quality of the statistical models, with a model with a lower BIC preferred. 
In the interaction model, this criterion was used to determine the appropriate value of $K$, the BIC of the Gaussian mixture models with different $K$ can be seen in Fig. \ref{fig:mesh5}. The BIC continues to decrease with an increase in number of components, with little change in rate (less than 10\%) when the number of components is larger than 10. Considering the computing time and the complexity of the models, $K$ = 10 was selected when generating the multivariate Gaussian mixture model.

%A BIC analysis can help to decide $K$ (number of components). The model with the lower BIC is preferred. The analysis result is shown in Fig. \ref{fig:mesh5}. As the figure indicates, BIC continues to decrease with the increase of the number of components, and the change rate is less than 10\% when the number of components is larger than 10, which means there will not be a big improvement to continue to raise $K$ after $K=10$. Considering the time costs of computing, $K = 10$ is a proper choice in this study to fit the data.

An interaction model was then created by fitting a 10-component truncated multivariate Gaussian mixture model to the collected naturalistic data using the algorithms above.
\subsection{Model Display and Utilization}

\begin{figure}[!h]
  \centering
  \subfigure[Raw data]{\includegraphics[width=0.23\textwidth]{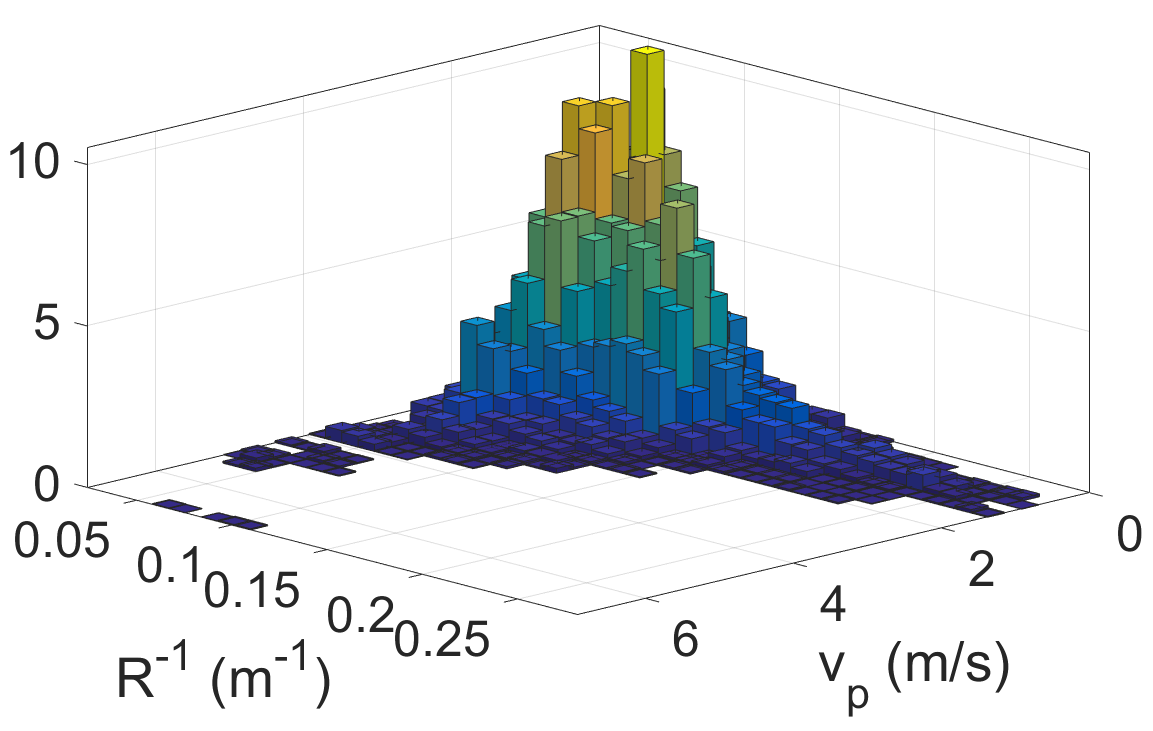}}
	\hfill
  \subfigure[Model distribution]{\includegraphics[width=0.23\textwidth]{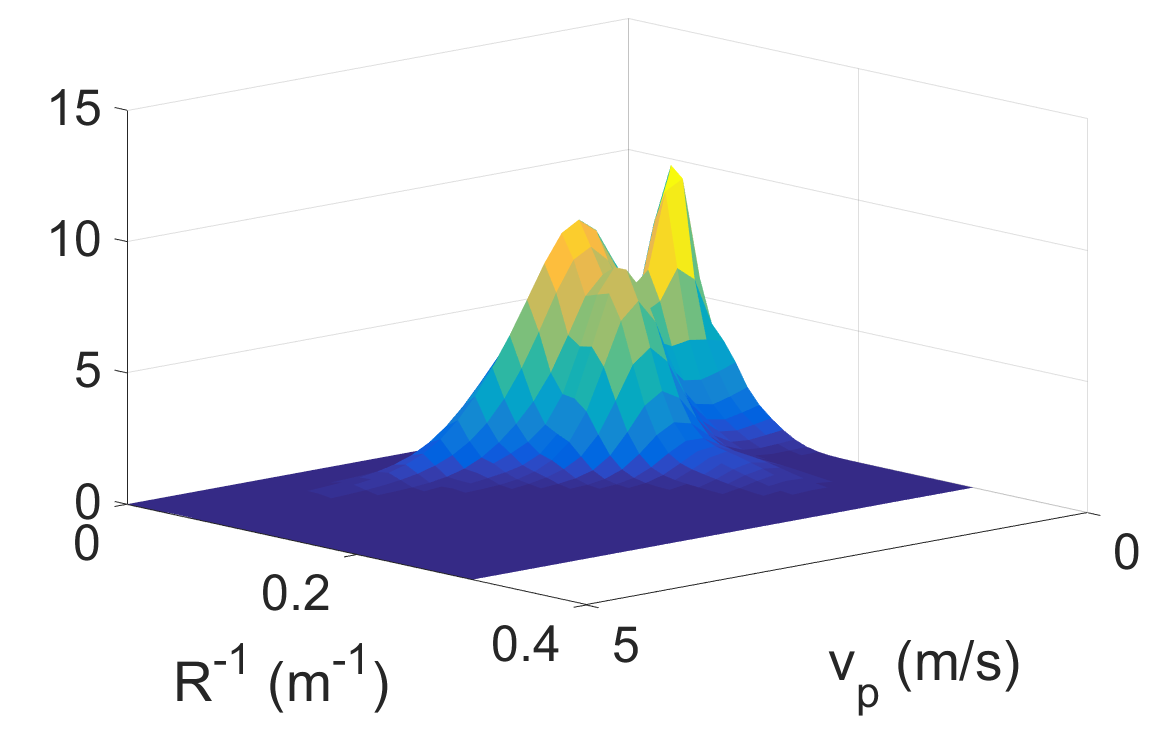}}
  \subfigure[Raw data]{\includegraphics[width=0.23\textwidth]{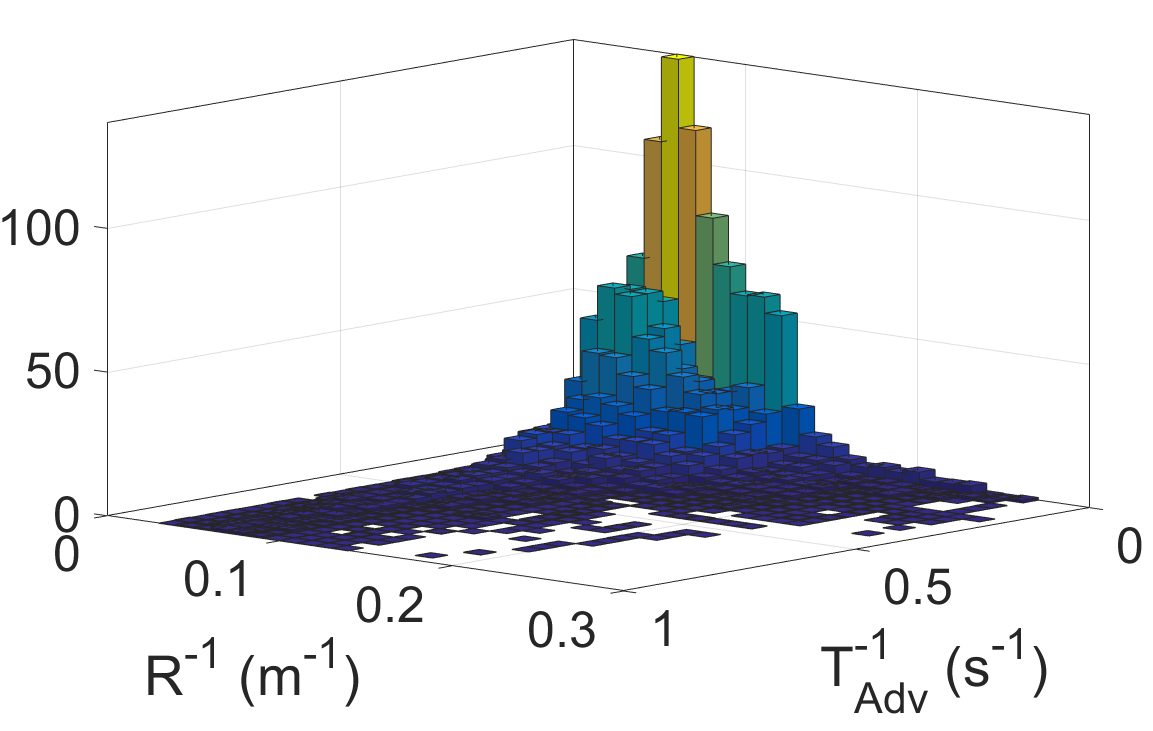}}
  \hfill
  \subfigure[Model distribution]{\includegraphics[width=0.23\textwidth]{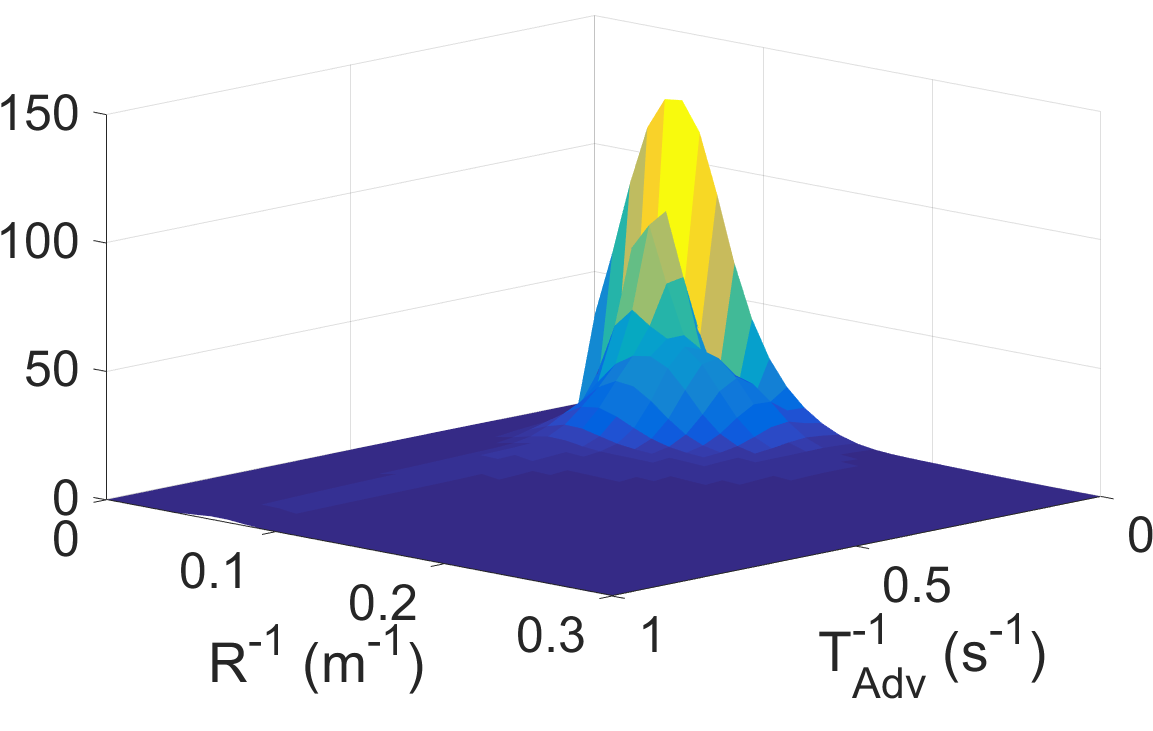}}
   \subfigure[Raw data]{\includegraphics[width=0.23\textwidth]{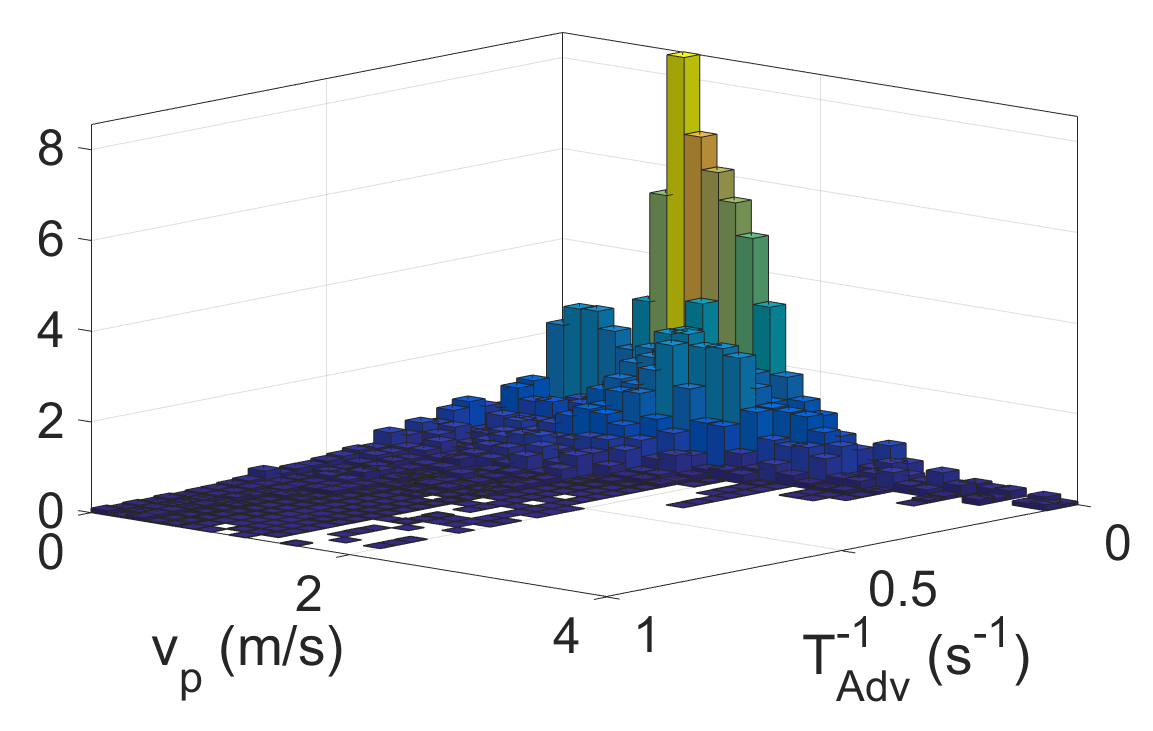}}
  \hfill
  \subfigure[Model distribution]{\includegraphics[width=0.23\textwidth]{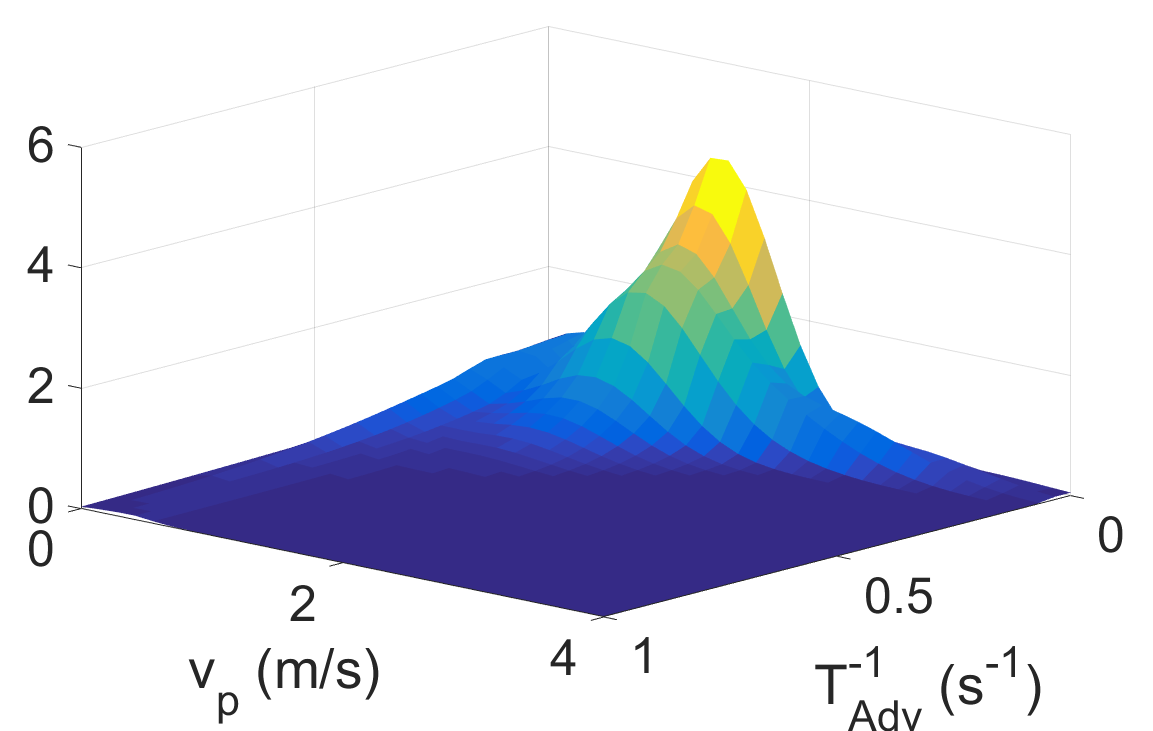}}
  \caption{Comparison of raw data and Gaussian mixture model}
  \label{fig:mesh6}
\end{figure}

Since there were four variables (i.e., \{$R^{-1}$, $v$, $v_p$, $T_{Adv}^{-1}$\}) in the Gaussian mixture model, the distribution of the interaction model was a 4-D function, which cannot be shown effectively on a flat piece of paper.
We projected the function to three 2-D functions to illustrate the probability distribution, and then compared them with the raw data collected. The results are shown in Fig. \ref{fig:mesh6}.

The stochastic interaction model developed here can help to simulate the walking speed of pedestrians. When encountering an unsignalized crossing, the pedestrian will decide the appropriate walking speed to cross the street, depending on the speed and position of the oncoming vehicle; that is, the distribution of pedestrian walking speed ($v_p$) is a conditional distribution of the multivariate Gaussian mixture model. By calculating this conditional distribution, the walking speed of pedestrians ($v_p$) can be simulated. 

For a Gaussian distribution, if $y=(y^T_m,y^T_o)^T$ is with mean $\mu$ and covariance $\Sigma$, then the conditional distribution of $y_m|y_o$ is also normally distributed with mean and covariance:
\begin{equation}
\mu_{m|o} = \mu_m+\Sigma_{m,o}\Sigma_{o,o}^{-1}(y_o-\mu_o),
\label{a1}
\end{equation}
\begin{equation}
\Sigma_{m|o}=\Sigma_{m,m}-\Sigma_{m,o}\Sigma_{o,o}^{-1}\Sigma_{o,m}.
\label{a2}
\end{equation}
To calculate the conditional distribution of a mixture Gaussian model, first calculate the conditional distribution of each component, then normalize all components and set new weights.

\begin{figure}[!h]
  \centering
  \includegraphics[width=0.5\textwidth]{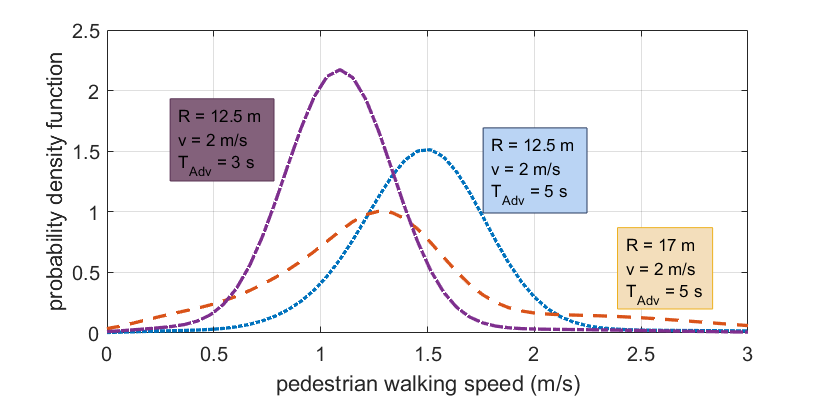}
  \caption{Probability density functions of pedestrian walking speed under different conditions}
  \label{fig:mesh9}
\end{figure}

If we set $y_m=v_p$ and $y_o = \{R^{-1}, v, T_{Adv}^{-1}\}$, then the distribution function of $v_p$ under given conditions can be calculated.
Examples of distribution functions under different conditions are shown in Fig. \ref{fig:mesh9}. 
We can generate a random $v_p$ from this distribution and take it as the walking speed of a pedestrian under the given conditions for the simulation and evaluation. 

%\newpage

\begin{figure}[h!]
  \centering
  \includegraphics[width=0.5\textwidth]{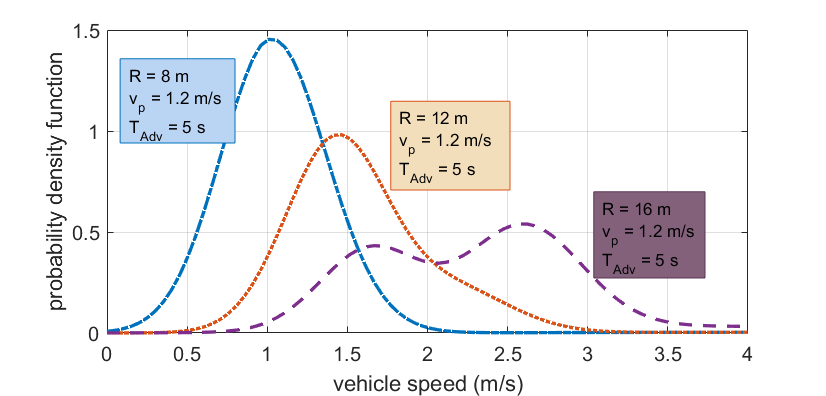}
  \caption{Probability density functions of vehicle speed under different conditions}
  \label{fig:mesh8}
\end{figure}

The movement of vehicles can also be simulated using the same method by calculating the conditional distribution. To do that, we set $y_m = v$ and $y_o = \{R^{-1}, v_p, T_{Adv}^{-1}\}$, and then the distribution function of $v$ can be calculated. Examples of distribution functions under different conditions are shown in Fig. \ref{fig:mesh8}. It is clear that, in this model, the vehicle is likely to have a lower speed when the distance is shorter, which makes sense considering traffic safety. The speed with the highest probability might be a good choice for setting the desired speed of the vehicle under given conditions when designing driving strategy.

\section{Simulation}

%Simulation was used to show the procedures used to evaluate the driving strategy of automated vehicles. The automated vehicle competed with the experienced human drivers for a better performance when encountering unsignalized crossings in the simulation. To achieve that, the driving strategy of automated vehicles should be neither too aggressive endangering traffic safety nor too conservative leading to a waste of time. With an increase in aggressiveness of the automated vehicle, less time will be needed to pass the intersection, resulting in a rise in the crash rate. As shown in Fig. \ref{fig:ag}, the aggressiveness needs to be within the allowed interval when the automated vehicle has a passing time less than $\mu_0$ and a crash rate less than $\kappa_0$.

\begin{figure}[!h]
  \centering
  \includegraphics[width=0.45\textwidth]{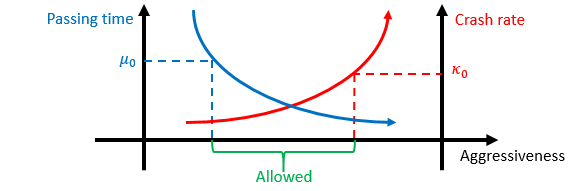}
  \caption{Influence of the vehicle's aggressiveness}
  \label{fig:ag}
\end{figure}

A simulation is used to show how to evaluate the driving strategy of automated vehicles. The automated vehicle is evaluated against the human drivers for better performance when encountering unsignalized crossings in the simulation. We expect that the driving strategy of automated vehicles be neither too aggressive (endangering traffic safety) nor too conservative (leading to traffic congestion). With an increase in aggressiveness of the automated vehicle, less time is then needed to pass through the intersection, resulting in a possible rise in the crash rate. As shown in Fig. \ref{fig:ag}, the aggressiveness needs to be within the allowed interval when the automated vehicle has a passing time less of than $\mu_0$ and a denied crash rate less than $\kappa_0$.

\subsection{Procedures for the Evaluation Experiment}
%PreScan software will be used for the simulation. The experiment will take place at an unsignalized crossing (Fig. \ref{fig:mesh10}). At first, the vehicle will come from a long distance at a constant speed $v_0$. When it is close, at a certain distance $R_0$, pedestrians will start to be generated to go across the street, and the vehicle will interact with pedestrians and try to pass through the intersection without any crashes. Each experiment will be done twice. The first time, the automated vehicle will be tested to interact with the pedestrians; the second time, a vehicle driven by an experienced human driver will replace the automated vehicle and interact with the same pedestrian, and the experiment data will be recorded as a reference for the evaluation. In this study, we simulated 50 times of experiments. The passing strategies of pedestrians, automated vehicles and human drivers simulated in this section will be described below separately. The results will be analyzed at the end of this section.

\begin{figure}[!h]
  \centering
  \includegraphics[width=0.5\textwidth]{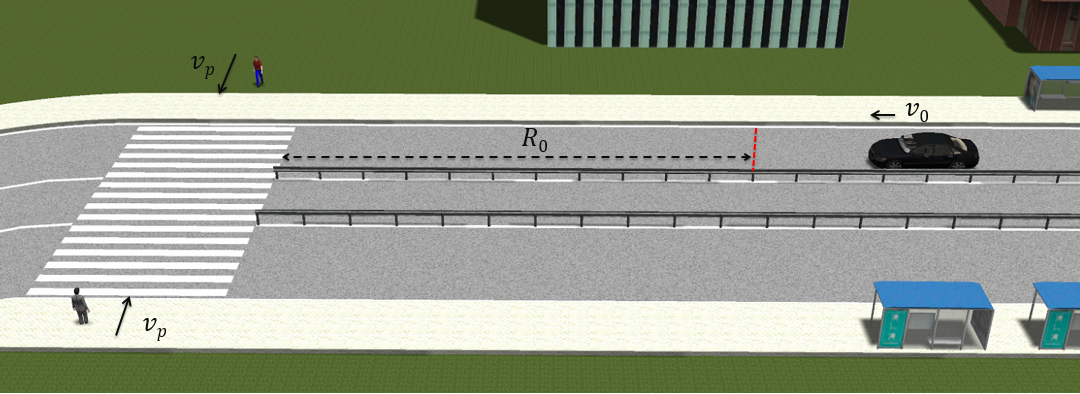}
  \caption{Environment of the simulation. When the vehicle is at a certain distance ($R = R_0$), pedestrians will start to arrive at the unsignalized crossing.}
  \label{fig:mesh10}
\end{figure}

PreScan\textsuperscript{\textregistered} is used for the simulation. The experiment takes place at an unsignalized crossing (Fig. \ref{fig:mesh10}). At first, the vehicle comes from a long distance at a constant speed $v_0$. When it is close, at a certain distance $R_0$, pedestrians will start to be generated to go across the street, and the vehicle will interact with pedestrians and try to pass through the intersection without any crashes. Each experiment is done twice. The first time, the automated vehicle is tested to interact with the pedestrians; the second time, the human driver passing strategy is applied to interact with the same pedestrian which is recorded as a reference for the evaluation. In this study, the simulation is run 50 times. The passing strategies of pedestrians, automated vehicles and human drivers simulated in this section are described below separately. The results are analyzed at the end of this section.

%The passing strategy of automated vehicles should be neither too aggressive nor too conservative. The former situation will endanger traffic safety, while the latter will lead to wasting of time and traffic congestion. At the same time, the passing behavior of the automated vehicles should be stable. 

%Thus three parameters were defined to evaluate the passing strategy

% \subsection{sensors}
% Antenna sensors will be installed on the vehicles and pedestrians, so that they can get needed information from each other, including their position and velocity, which will be needed to decide their passing strategy. A TIS sensor (Radar/Laser) will also be installed on the vehicle for the convenience of observation and analysis. 

\subsection{Pedestrian Crossing Strategy}
\begin{figure}[!h]
  \centering
  \includegraphics[width=0.5\textwidth]{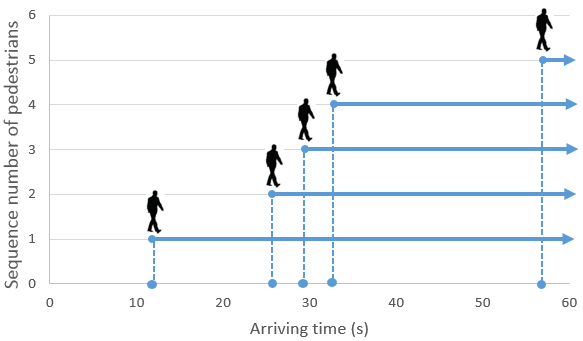}
  \caption{Arriving times of pedestrians within 60 seconds}
  \label{fig:mesh11}
\end{figure}

Pedestrians are made to cross the street from both sides. When the vehicle is at a certain distance ($R = R_0$), pedestrians will start to arrive at the unsignalized crossing. Each pedestrian has an arriving time when they will be put on the side of the street and start to cross. The arriving time of the pedestrians obeys the Poisson process:
\begin{equation}
P\{ N(t)=n \} =  \frac{(\lambda t)^n}{n!} e^{(-\lambda t)}.
\end{equation}
Parameter $\lambda$ can be set to different values depending on the density of pedestrian flow, which varies greatly depending on time and location. For reference, pedestrian flow in the campus area during peak hours is around 250 peds/hr \cite{Schroeder2014}.
Based on this distribution, an arriving time example was generated and is shown in Fig. \ref{fig:mesh11}. Five pedestrians are scheduled to arrive at this crossing within 60 seconds.

Though the pedestrians' arriving times are random, their walking speeds are usually not.
Each pedestrian has a constant walking speed that is calculated by the interaction model depending on the distance and speed of the oncoming vehicle when the pedestrian arrives, which is similar to real world process, in which most pedestrians will look at the oncoming vehicle and decide what to do before moving.

%One wall will block the pedestrians before they are at a certain distance ($L_0$) to the center line of the street so that there will be a clear time when the vehicle starts to detect the pedestrian with its TIS sensor. The walls will not block the communication of Antenna sensors.

\subsection{Automated Vehicle's Passing Strategy}

The driving strategy evaluated in the simulation is the Soft-Yield model proposed by Schroeder et al. \cite{Schroeder2014} which provides vehicle trajectories when facing one pedestrian.
%However, very few models have been developed currently, especially in scenarios where the vehicle is facing multiple pedestrians. 

\begin{figure}[!h]
  \centering
  \includegraphics[width=0.3\textwidth]{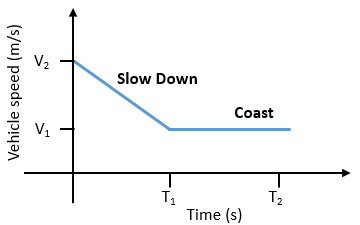}
  \caption{Driving strategy of the automated vehicle when encountering an unsignalized crossing (developed by Schroeder et al. \cite{Schroeder2014})}
  \label{fig:mesh12}
\end{figure}

The model was developed based on GPS data from the instrumented vehicle study. The generalized vehicle distance-speed and time-speed models are shown in Fig. \ref{fig:mesh12}. When approaching the unsignalized crossing, the vehicle has a constant deceleration for time length $T_1$ and then starts to coast to the crosswalk.
%To our best knowledge, there is currently only one model available that provide vehicle trajectories proposed by Schroeder et al. in 2014\cite{Schroeder2014}. The generalized vehicle distance-speed and time-speed models are shown in Fig. 11. 

Based on a regression analysis, vehicle acceleration $a$ is set to
\begin{equation}
%\begin{aligned}
a = p_1 + p_2v + p_3R,
%\end{aligned}
\end{equation}
where $p_1 = 0.0169$, $p_2 = -0.13986$, $p_3 = 0.010115$. $v$ and $R$ are the vehicle's information when the vehicle perceives the pedestrian and makes its decision. 

Deceleration time $T_1$ is calculated as follows:
\begin{equation}
t_L = \frac{L_0}{v_p},
\end{equation}

\begin{equation}
T_1 = t_L - \sqrt{t_L^2-\frac{2(R-v*t_L)}{a}},
% T1 = VehTime - 
% \sqrt{(VehTime)^2-\frac{2*(Distance - Speed*VehTime)}{Decel}}
\end{equation}
where $L_0$ is the length of the crossing path, and $t_L$ is the time needed for the pedestrian to complete the crossing.
%In addition, this model is feasible when there is only one pedestrian. There are currently no driver models developed for multiple pedestrians.

\subsection{Human Driver's Passing Strategy}
%% our own strategy
%Ideally, humans should be driving to serve as a reference for evaluating the automated vehicles. For simulation purposes, however, 
A human driver's passing strategy was developed by studying the interaction model based on the naturalistic traffic data. 
When encountering an unsignalized crossing, the strategy of the simulated human driver is set to adjust the vehicle's speed if a pedestrian is detected. The desired speed $v_d$ for adjusting is calculated based on the position and velocity of the pedestrian in the current state. Thus, the desired acceleration of the vehicle is
\begin{equation}
a_d = \frac{v_d-v}{\Delta t},
\end{equation}
where $\Delta t$ is the time interval between two updates of the desired speed. In this study, we set $\Delta t$ = 1 s.

Each vehicle, of course, has a maximum limit of acceleration $a_m$. Thus, if $a_d>a_m$, the acceleration of the vehicle $a$ will be set to $a=a_m$. Otherwise, the acceleration will be set to just $a=a_d$.

When no pedestrian is present within the radar's detection range, or the pedestrians have left the crossing, the vehicle will accelerate to $v_0$ at a constant speed of $a_0$ = 1 m/s. 

%%
% The whole experiment will process 3 stages and 2 transition points:

% stage 1($0<t<t_1$):
% The automated vehicle is coming to the crossing from a distance with a constant speed. The pedestrians stands still. The vehicle is not able to see the pedestrian.

% Transition point 1($t = t_1$):
% When the vehicle is at a certain distance($R = R_0$), the first pedestrian will start to walk across the street at a constant speed that is calculated with our model. The arriving time of other pedestrians obeys Poisson arrival distribution:

% \begin{equation}
% P\{ N(t)=n \} =  \frac{(\lambda t)^n}{n!} e^{(-\lambda t)}
% \end{equation}

% The parameter($\lambda$) of the Poisson distribution will be set depending on the density of the traffic flow.

% stage 2($t_1 < t < t_2$):
% The vehicle and the pedestrian continue to move at a constant speed. The vehicle is still not able to see the pedestrian.

% Transition point 2($t = t_2$):
% When the pedestrian is at a certain distance($L_0$) to the center line of the street, the vehicle starts to see the pedestrian.

% stage 3($t_2 < t$):
% The vehicle adjusts its speed with passing strategies, and decides whether to yield to the pedestrians. Experiment data and result will be observed and collected.

\subsection{Simulation Results}
All the above procedures are simulated using PreScan\textsuperscript{\textregistered}. The parameters set for the simulation is shown in Table \ref{table2}. One pedestrian is generated in each experiment. 

\begin{table}[!ht]
\caption{Parameters in the simulation}
\begin{center}
\begin{tabular}{ccc}
\hline
\hline
symbol & unit & value \\
\hline
$R_0$ & m & 30\\
$L_0$ & m & 9\\
$v_0$ & m/s & 5\\
\hline
\end{tabular}
\end{center}
\label{table2}
\end{table}
%We set $R_0$ = 30 m and $L_0$ = 9 m. At first, the vehicle runs at a constant speed of $v_0$ = 5 m/s. 

% The parameters used for the evaluation are defined in Table \ref{table2}.

% \begin{table}[!ht]
% \caption{List of parameters in the evaluation procedure}
% \begin{center}
% \begin{tabular}{ccc}
% \hline
% \hline
% symbol & unit & value \\
% \hline
% $N$ & - & number of experiments\\
% $t_{a}^i$ & s & time used by the automated vehicle in experiment $i$\\
% %\hline
% $t_h^i$ & s & time used by the human driver in experiment $i$\\
% %\hline
% $\tau_i$ & - & $t_{a}^i/t_h^i$, ratio of used time in experiment $i$\\
% %\hline
% $\mu$ & - &mean value of $\{\tau_i\}$ \\
% %\hline
% $\sigma$ & - &standard deviation of $\{\tau_i\}$\\
% $c_v$ & - &coefficient of variation\\
% %\hline
% $n_{crash}$ & - &number of crashes happened \\
% %\hline
% $\kappa$ & - &$n_{crash}/N$, frequency of crashes\\
% \hline
% \end{tabular}
% \end{center}
% \label{table2}
% \end{table}

Each experiment is done twice. The first time, the automated vehicle's strategy is tested to interact with the pedestrians generated based on the Poisson process; the second time, the human driver's strategy is tested to interact with the same pedestrian. The time used and the safety of the pedestrian is recorded. We evaluate the aggressiveness of the automated vehicle by analyzing its results compared to those of the human driver strategy.

\begin{figure}[!h]
  \centering
  \includegraphics[width=0.5\textwidth]{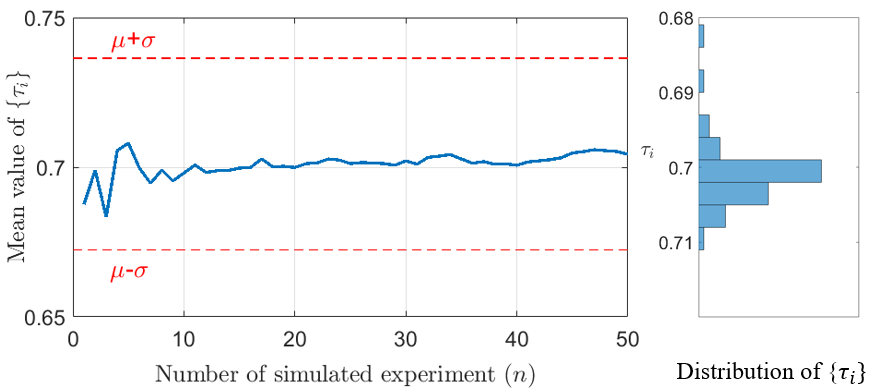}
  \caption{Mean value and distribution of $\{\tau_i\}$ (the ratio of time used in experiments $i$). As the number of simulated experiment increases, the mean value of $\{\tau_i\}$ approaches a limit value.}
  \label{fig:mesh133}
\end{figure}

Assume $t_{a}^i$ and $t_h^i$ are the time used by the automated vehicle and the human driver for passing through the unsignalized crossing in experiment $i$, respectively. Then, $\tau_i = t_{a}^i/t_h^i$ indicates the efficiency of the automated vehicle compared to the human driver.
After $N$ = 50 experiments, the mean value and distribution of $\{\tau_i\}$ is shown in Fig. \ref{fig:mesh133}. As $n$ increases, the mean value of $\{\tau_i\}$ approaches a limit value, which means that the time used by the automated vehicle to pass through the intersection is stable.

%\newpage
$\mu$ (The mean value of $\{\tau_i\}$), $c_v$ (the coefficient of variation of $\{\tau_i\}$) and $\kappa$ (the crash rate) are parameters to indicate the passing strategy's efficiency, stability and safety, respectively. They are calculated to evaluate the automated vehicle based on the experimental results:

\begin{equation}
\mu = \frac{\sum\tau_i}{N}=0.7044,
\end{equation}
% \begin{equation}
% \sigma = \sqrt[]{\frac{1}{N}\sum_{i=1}^{N}(\tau_i-\mu)^2}=0.0320
% \end{equation}
\begin{equation}
c_v = \frac{\sigma}{\mu} = \frac{\sqrt[]{\frac{1}{N}\sum_{i=1}^{N}(\tau_i-\mu)^2}}{\mu} = 4.54\%，
\end{equation}
\begin{equation}
\kappa = \frac{n_{crash}}{N}=0.
\end{equation}

It turns out that the passing strategy of the automated vehicle is more efficient than that of a human driver, taking only about 70\% of the time a human driver uses to pass the unsignalized crossing. It is also quite stable since $c_v$ is under 5\%. More importantly, no crash occurred during the simulation, meaning the strategy is also safe under ideal conditions.

\section{Conclusion and Discussion}

%We proposed an evaluation method of automated vehicles' interactions with pedestrians at unsignalized crossings, including a series of algorithms to process the data collected. We simulated the interaction by learning the naturalistic data and apply it to test the passing strategies of automated vehicles. 

%In this research, we did a series of work including data collection, interaction modeling, and designing of an evaluation method for testing the safety and feasibility of automated vehicles' passing strategy at unsignalized crossings. One Soft-Yield driver model was evaluated in the experiment.

In this paper, we proposes a method to evaluate the passing strategy of automated vehicles at unsignalized crossings. A stochastic interaction model is developed to predict pedestrian movement. A simulation is conducted to demonstrate the evaluation procedure. The automated vehicle is evaluated against the human drivers for better performance in terms of the time used and the crash rate. One Soft-Yield driver model is evaluated as an example. The simulation results indicate that this driving strategy is more efficient than that of a human driver that is modeled by naturalistic data collected in Ann Arbor.

The pedestrians in our evaluation experiments comes from both sides of the crossing. Their arriving times obey the Poisson process; the density of pedestrian flow can be adjusted by setting different values of parameters. The pedestrians in the simulation will have a look at the oncoming vehicle and then decide on their walking speed before moving, similar to the process in the real world. 
However, a more detailed pedestrian model remains to be developed, and more behaviors and features can be taken into account in future work.

\section*{Disclaimers}

This work was funded in part by the University of Michigan Mobility Transformation Center Denso Tailor project. The findings and conclusions in the report are those of the authors and do not necessarily represent the views of the MTC or Denso.

%However, very few driver models have been developed currently, especially in scenarios when the vehicle is facing multiple pedestrians. It is possible to build a new driver model based on the data we collected.

\bibliographystyle{IEEEtran}
\bibliography{Mendeley_Baiming_Ding.bib}

\end{document}